# Magnon sensing of NO, $NO_2$ and $NH_3$ gas capture on CrSBr monolayer

Gonzalo Rivero-Carracedo[†], Andrey Rybakov[†] and José J. Baldoví[†,*]

[†]Instituto de Ciencia Molecular, Universitat de València, Catedrático José Beltrán 2, 46980 Paterna, Spain. E-mail: j.jaime.baldovi@uv.es

**Abstract**

Air pollution and greenhouse emissions are a significant problem across various sectors, urging the need for advanced technologies to detect and capture harmful gases. In recent years, two-dimensional (2D) materials have attracted an increasing attention due to their large surface-to-volume ratio and reactivity. Herein, we investigate the potential of single-layer CrSBr for gas sensing and capturing by means of first-principles calculations. We explore the adsorption behaviour of different pollutant gases ($H_2S$, $NH_3$, NO, $NO_2$, CO and $CO_2$) on this 2D ferromagnet and the impact of intrinsic defects on its magnetic properties. Interestingly, we find that Br vacancies enhance the adsorption of $NH_3$, NO and $NO_2$ and induces a selective frequency shift on the magnon dispersion. This work motivates the creation of novel magnonic gas sensing devices based on 2D van der Waals magnetic materials.

**Introduction**

Air pollution and greenhouse gas emissions are a major concern in various fields, including industry, public health, or agriculture, which call for novel technologies to detect and trap harmful and toxic gases.[1] Recently, two dimensional (2D) materials have gained an increasing interest in gas sensing and capturing devices owing to their larger surface to volume ratio and reactive surface,[2–4] which present an advantage over the main limitations of the widely investigated metal oxide semiconductors (MOS).[5,6] In this context, graphene-based gas sensors have already shown their ability to detect concentrations as low as 1 ppb, exhibiting pronounced sensitivities of many orders of magnitude higher than previous devices.[7,8] Other 2D inorganic materials such as silicene, $MoS_2$, $WS_2$ or GaN (among others) have also been studied showing exceptional performance, which opens new avenues for the design and fabrication of novel devices with improved stability and sensitivity.[9–11] Besides, their easy tunability and van der Waals (vdW) nature allow to design structures based on atom doping,[12] vacancy engineering,[13] intercalation,[14] and creation of heterostructures.[15]

An emerging class of gas sensing/capturing devices is based on their magnetic response. These so-called magnetic gas sensors offer a series of advantages when compared to the electrical-based ones, namely: (i) no need of electric contacts for gas detection, (ii) faster responses, and (iii) tunability of the working temperature by choosing magnetic materials with different Curie temperature ($T_C$).[16] Indeed, magnetic gas sensors that rely on the measurement of magnetostatic spin waves have also been developed using the high-$T_C$ Yttrium Iron Garnet (YIG). These systems are called magnonic gas sensors and have shown operative room temperature, high sensitivity, short response time and good reproducibility.[17,18] Although magnons have been experimentally detected in 2D vdW magnetic materials, the development of magnonic gas sensors using 2D materials –even from a theoretical point of view– is still unexplored.

In this work, we select the air-stable 2D semiconducting ferromagnet CrSBr as a model system, which exhibits high energy magnons and $T_C$~146 K.[19–24] We investigate the capability of this material to sense and/or capture common air pollutants such as $H_2S$, $NH_3$, NO, $NO_2$, CO and $CO_2$ using first-principles calculations. In addition, we explore how the adsorption behaviour, magnetic properties and magnon dispersion can be modified via intrinsic defects, namely Br and S vacancies. Our findings show that Br vacancies on single-layer CrSBr can trap $NH_3$, NO and $NO_2$, thus inducing a selective modification of the magnon frequencies that allows to selectively detect them. Our results open new avenues for the design and fabrication of 2D magnonic devices for gas capturing and sensing at the limit of miniaturization.

**Results**

Bulk CrSBr is a vdW layered material that possesses an orthorhombic crystal structure with a Pmmn space group.[25] In this material, the layers are weakly interacting via vdW forces, allowing the isolation of single-layer flakes via exfoliation. In each layer, the Cr atoms are arranged in a distorted octahedral coordination environment. They are linked to their neighbouring Cr atoms by S and Br bridges along the *a*-axis. Conversely, Cr atoms are only connected along the *b* axis through S atoms (see **Figure 1**), resulting in decoupled quasi-one-dimensional chains as proved by conductivity measurements.[26] According to our calculations, the relaxed lattice parameters are *a*= 3.545 and *b*= 4.735

Å for the monolayer, which are in good agreement with the reported cell dimensions in previous works.[25,27]

First, we design hybrid heterostructures formed by a single layer of CrSBr and targeted gas species ($H_2S$, $NH_3$, $NO$, $NO_2$, $CO$ and $CO_2$) to explore the adsorption behaviour and stability of these systems. We performed first-principles simulations based on density functional theory (DFT) considering five different adsorption sites, namely top of $Cr_1$, $Cr_2$, S, Br and hollow (**Figure 1b**). This allows us to determine the most favourable geometries of these molecules on the substrate according to the calculated adsorption energies ($E_{ads}$). For each molecule, we consider two initial opposite orientations, e.g. for the $H_2S$ molecule, the H atoms are pointing to/away from the slab. The optimized geometries are shown in **Figure 1c-h**. The preferential site, $E_{ads}$, and molecule-surface distances are summarized in **Table 1**. As we can observe, all optimized structures show negative $E_{ads}$, which indicates that the molecular adsorption in every system is an exothermic process and the formation of the heterostructures is thermodynamically favoured. According to our results, the adsorption capacity of the six molecules on CrSBr monolayer decreases in the order of $CO_2 > NO > NO_2 > H_2S > NH_3 > CO$. Regarding the equilibrium distance, in all cases is around 3 Å, except NO, which remains closer at 1.87 Å, where the N atom is pointing towards the substrate. All these values are compatible with a physisorption process in which no new chemical bonds are formed, and the main interactions are vdW forces. Moreover, the adsorption of the gas molecules does not induce a significant change in the bond lengths of the CrSBr substrate.

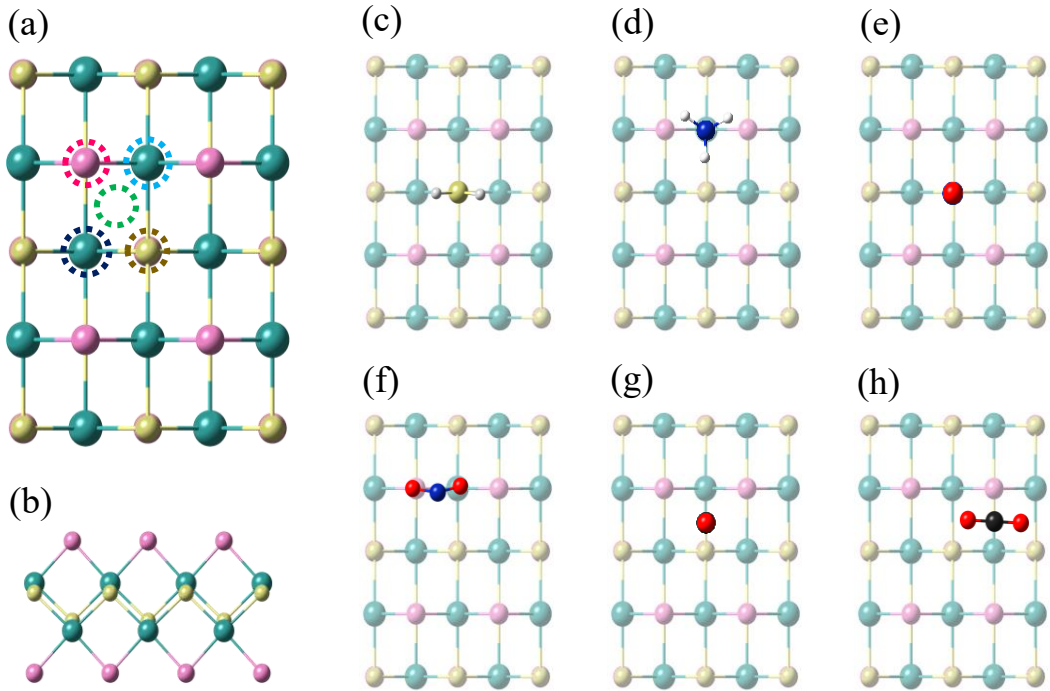

**Figure 1**. (a) Top view of monolayer CrSBr indicating the studied adsorption sites, namely $Cr_1$-top (dark blue circle), $Cr_2$-top, (light blue circle), S-top (golden circle), Br-top (pink circle) and Hollow-top (green circle). (b) Side view of monolayer CrSBr. (c-h) Optimized geometries of the gas/2D hybrid heterostructures formed by $H_2S$, $NH_3$, $NO$, $NO_2$, $CO$ and $CO_2$, respectively, over single-layer CrSBr.

**Table 1**. Preferential site, $E_{ads}$, and equilibrium distance for the different gas species over CrSBr. Note that 'Distance' is the distance between the upper atom of the substrate and the lower atom of the molecule.

| Molecule | Site | $E_{ads}$ (eV) | Distance (Å) |
|---|---|---|---|
| $H_2S$ | S-top | -0.109 | 3.013 |
| $NH_3$ | $Cr_2$-top | -0.095 | 2.953 |
| NO | S-top | -0.560 | 1.870 |
| $NO_2$ | Hollow | -0.404 | 2.821 |
| CO | S-top | -0.049 | 3.009 |
| $CO_2$ | $Cr_1$-top | -0.838 | 3.002 |

To investigate the charge redistribution between the adsorbates and the 2D material, we calculate the charge density difference (CDD) of the gas/CrSBr heterostructures. The plots are shown in **Figure 2,** where the yellow and blue lobes represent charge accumulation and depletion, respectively. Note that we set the same isosurface value of $5 \cdot 10^{-4}$ eV·Å$^{-3}$ in all the systems for comparison. Charge polarization is only observed for $NH_3$, NO and $NO_2$, being the most significant for NO, which is the closest molecule to the surface. The study of the charge transfer mechanism by means of a Bader analysis reveals the donor behaviour of NO, which transfers 0.103e towards the substrate. This is mainly attributed to its relative orientation, where the lone electron pair of the N atom points towards the Br atoms of the substrate, thus permitting the donation of charge density. Such electron density is accepted by the Br ligands of the substrate due to their higher electronegativity. By contrast, the effects of the rest of the molecules are negligible, given that they are not able to transfer a significant amount of charge because they remain further away from the substrate. This is also consistent with the calculated band structures, where only NO has a remarkable effect on the electronic levels, showing n-type behaviour that pushes the Fermi energy towards the conduction band and adds two molecular localized states that lie close to the Fermi level (**Figure S5**).

Next, we study the adsorption behaviour of the gas molecules onto a defective CrSBr monolayer. For this purpose, we create single Br ($V_{Br}$-CrSBr) and single S ($V_S$-CrSBr) vacancies by removing such atoms from the surface where we consider a density of defects of $3.72 \cdot 10^{13}$ cm$^{-2}$. We study both Br and S vacancies because they are the most probable to form in the material from a thermodynamic point of view.[28] The optimized geometries of the gas molecules over $V_{Br}$-CrSBr and $V_S$-CrSBr are shown in **Figures S1 and** S2. Regarding the value of the $E_{ads}$ (**Figure 3**), $V_S$-CrSBr does not display a noticeable change with respect to pristine CrSBr. On the other hand, $E_{ads}$ for $V_{Br}$-CrSBr decreases dramatically (enhanced interaction) in the case of $NH_3$, NO and $NO_2$, indicating that the adsorption behaviour switches from physisorption to chemisorption for such molecules. In particular, $NH_3$, NO and $NO_2$ occupy the cavity left by the missing Br atom, where the N atom is forming a chemical bond with the Cr atom from the substrate.

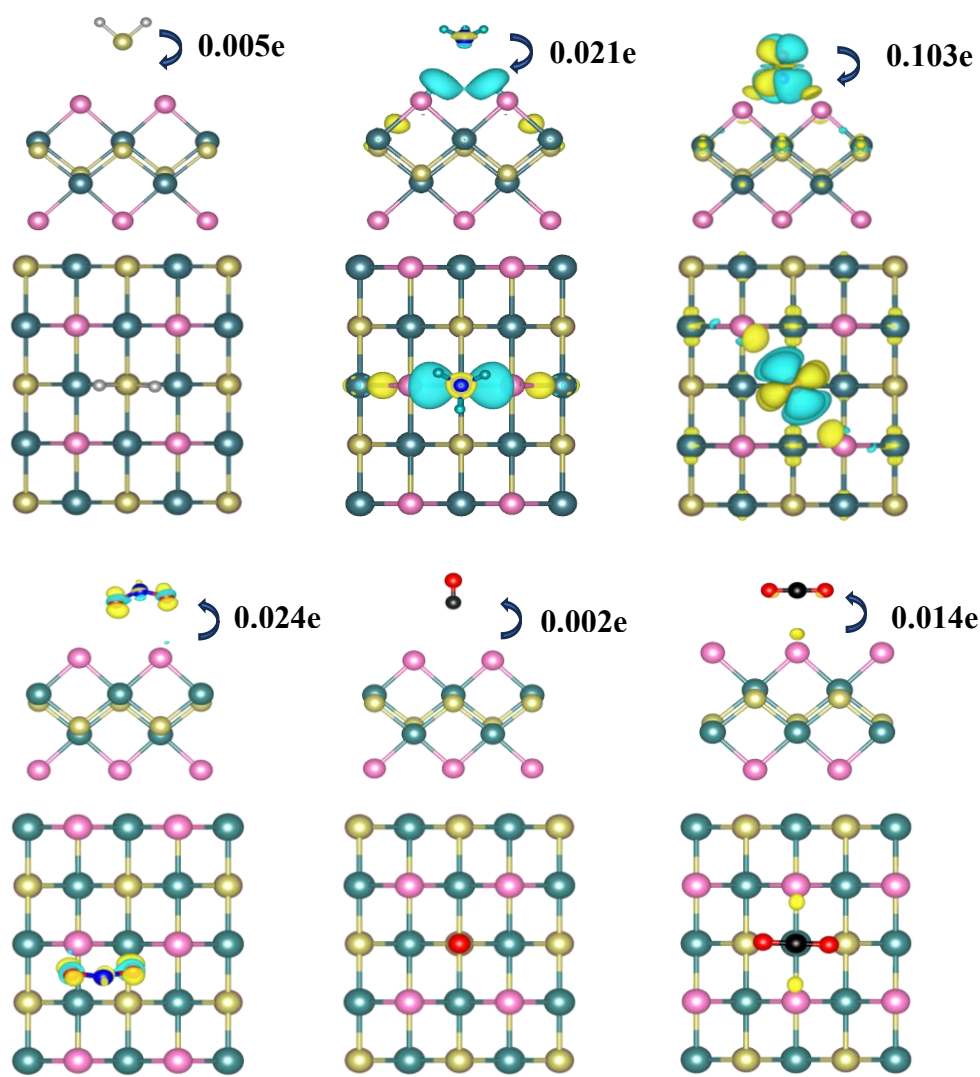

**Figure 2**. CDD plots for the gas/CrSBr hybrid heterostructures. Blue (yellow) lobes represent charge depletion (accumulation). For comparison, all isosurface values are set to $5\cdot 10^{-4}$ eV·Å$^{-3}$

The CDD plots (

**Figure 4**) for the gas/V$_{Br}$-CrSBr heterostructures illustrate the charge redistribution caused by molecular adsorption. Compared with the non-defected surface Figure **2**), electronic density lobes expand dramatically for the three chemisorbed molecules (NH$_3$, NO and NO$_2$). This enhancement on the electrostatic interaction strength is also consistent with the Bader analysis, where the electron transfer is notably increased between H$_2$S, NH$_3$, NO and NO$_2$ and the substrate, being 0.108e, -0.103e, 0.568e and 0.636e, respectively (**Figure S4**). In the case of NH$_3$, the donation of electron density can be attributed to the lone pair present in the N atom, which forms a chemical bond with Cr and thus acts as a ligand. By contrast, for NO and NO$_2$ we observe an acceptor behaviour, which is mainly due to the high electronegativity of the O atoms present in both molecules, which accept the electron density captured from the substrate. In the case of H$_2$S, it is not chemically occupying the cavity left but there is an increase of the charge

transfer which is accepted by the S atom. On the other hand, CDD for $V_S$-CrSBr (**Figure S3**) does not show a significant modification in terms of charge density distribution induced by the molecules compared to the adsorption over pristine CrSBr. In fact, S vacancies are much more inner with respect to Br counterparts, being impossible for the gas species to diffuse through the material to occupy such empty space.

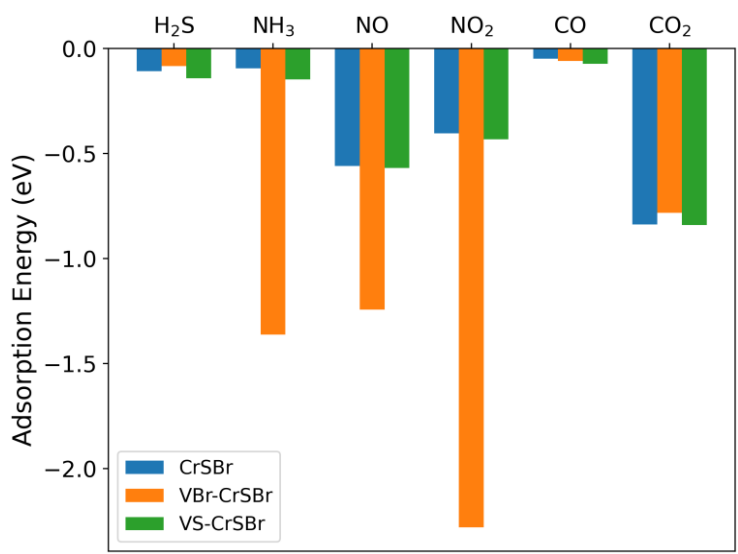

**Figure 3**. Calculated adsorption energy of the different gas species over CrSBr, $V_{Br}$-CrSBr and $V_S$-CrSBr

Then, we study the potential application of CrSBr monolayer in gas detection systems of the targeted gas molecules by calculating their corresponding recovery times ($\tau$). $\tau$ is the time required for a sensor to regain its initial sensitivity following exposure to a specific gas concentration. In this context, a good recovery time for a reusable gas sensor should be in the range of seconds to a few minutes. Our results suggest that pristine CrSBr, $V_{Br}$-CrSBr and $V_S$-CrSBr will have suitable performance in detecting $CO_2$, with respective recovery times of 147, 17 and 168 s (**Table S3**). By contrast, the rest of the physisorbed molecules show much lower values. On the other hand, the chemisorption of $NH_3$, NO and $NO_2$ on $V_{Br}$-CrSBr leads to a big enhancement of $\tau$ to very high orders of magnitude ($10^{26}$ in the case of $NO_2$), which indicates that these molecules are trapped by the substrate in an irreversible process, which prevents their use as gas sensors but allows to chemically trap such pollutant gases.

Subsequently, we investigate the effect of the deposition of the selected gas molecules on the magnetic properties of CrSBr. For the pristine system, the calculated magnetic moment is 2.73 $\mu_B$, in agreement with previous theoretical works.[29,30] According to our results, the molecules do not induce a significant modification of the magnetic moments of the substrate: the largest change is for NO, which results in an enhancement of $7.5 \cdot 10^{-3}$ $\mu_B$ (see **Figure S9** for details). In addition, we study the exchange interactions up to third-nearest neighbours, which are extracted by mapping total energies from different magnetic configurations (**Figure S12**) onto an isotropic classical Heisenberg Hamiltonian. The calculated exchange parameters of pristine CrSBr are $J_1$ = 2.94 meV,

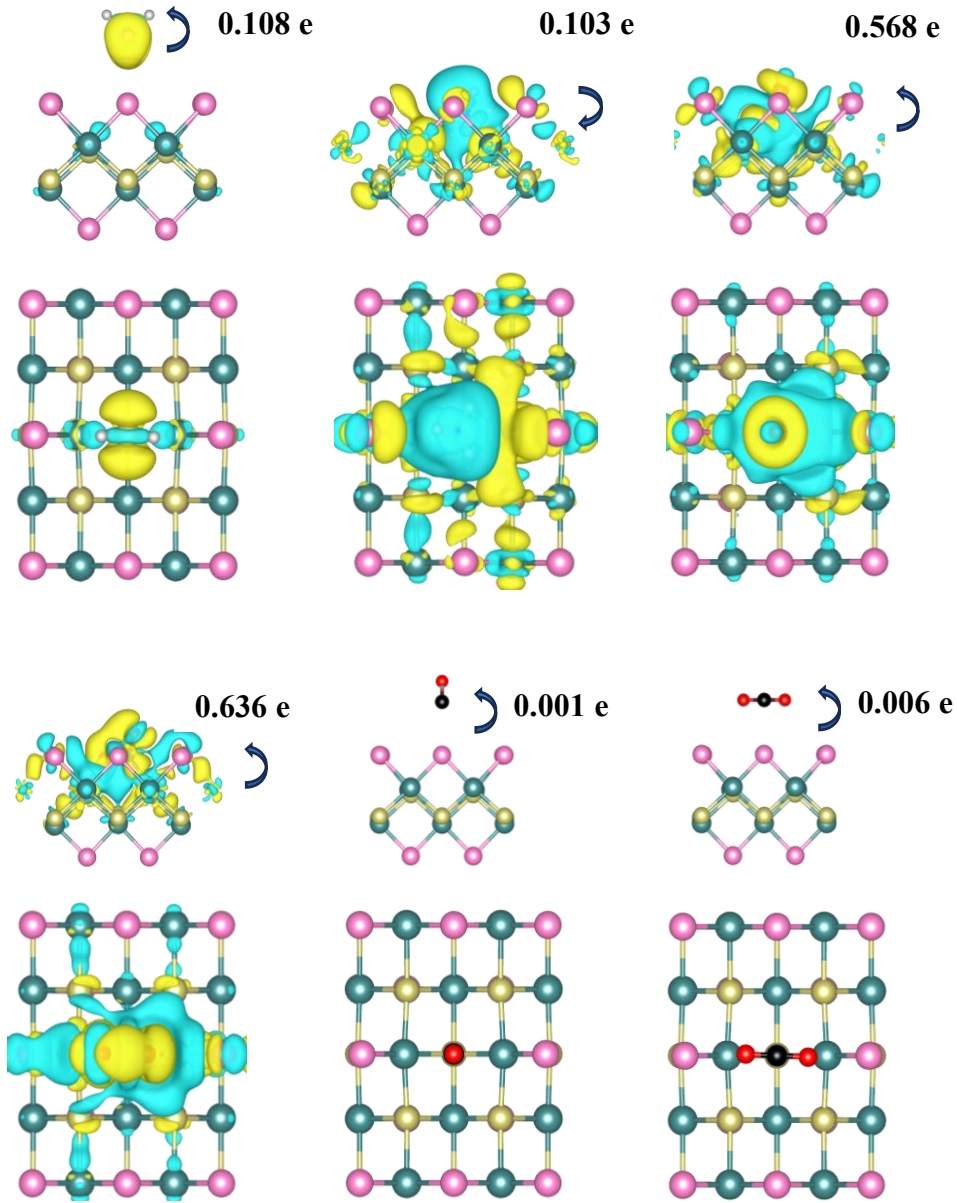

**Figure 4**. CDD plots for the gas/VBr-CrSBr hybrid heterostructures. Blue (yellow) lobes represent charge depletion (accumulation). For comparison, all isosurface values are set to $5 \cdot 10^{-4}$ eV·Å$^{-3}$

$J_2$ = 4.00 meV, and $J_3$ = 2.24 meV, which agree well with respect to the experimental values.[20] In this regard, the magnetic exchange parameters remain almost unaffected by the gas molecules, which is expected given that both structural and charge transfer effects are weak. Again, the only pronounced effect appears for NO, which increases $J_3$ up to 12% (**Figure S14**), which mediates the magnetic exchange in the easy magnetization axis of the material (b axis).

Regarding $V_{Br}$-CrSBr and $V_S$-CrSBr, we observe an enhancement of the magnetic moments of the Cr atoms that were previously bonded to the Br and S missing atoms, with a change of more than 10% with respect to the pristine substrate. This is attributed to the fact that both Br and S were mediating superexchange pathways (Cr-Br-Cr and Cr-

S-Cr) in which the magnetic moments were delocalized, being now localized in the Cr close to the vacancy (**Figure S8**). Regarding the magnetic coupling parameters, Br vacancies are mostly increasing $J_1$ (**Figure S13**), which is the exchange pathway in which Br atoms participate. In the case of S vacancies $J_3$ is noticeably increased given that this pathway is only mediated by S atoms. Then, we analyse the effects of the deposition of the molecules on the defected regions, where a substantial decrease of the Cr magnetic moments along the whole surface for the chemisorbed molecules ($NH_3$, $NO$ and $NO_2$) can be seen in the case of $V_{Br}$-CrSBr (**Figure S10**). Such modification is reflected in a substantial change in the exchange parameters, mainly decreasing $J_1$ and increasing $J_3$ (which are the ones mediated by Br), leaving $J_2$ almost unaltered. In the case of $V_S$-CrSBr, only CO is modifying the magnetic moments with a minor increase around 0.1 $\mu_B$ in the whole system, which does not imply a significant change in the magnetic exchange couplings.

Then, we calculate the magnon dispersion relation via the Holstein-Primakoff (HP) transformation[31] (**Figure 5a,b**). The HP boson expansion provides a reasonable description of the low energy magnetic excitations for the low temperature, where the long-range magnetic order is well established. Please note that this description imposes the condition of the low-temperature regime for the gas sensor. Because of the presence of two magnetic Cr atoms in the unit cell, we observe both acoustic (bottom branch) and optical (top branch) modes, being degenerated along the X-S and S-Y directions. On the other hand, we can observe the appearance of highly dispersive bands along the Γ-X and Γ-Y directions, which are directly related to the *a* and *b* crystallographic axes of CrSBr. The magnon dispersion upon molecular deposition on the non-defective substrate is moderately sensitive with the main shift produced by the adsorption of NO, which is due to the above-mentioned enhancement of $J_3$. Interestingly, the substrate that contains the Br vacancy is considerably more sensitive to the gas environment, as it was mentioned above. The most pronounced changes are caused by the presence of NO, $NO_2$ and $NH_3$ molecules. It can be noticed that the energy of the optical branch decreases along almost the whole k-path. In this case, such change is attributed to the lowering of the exchange parameter $J_1$, whose modification's magnitude is bigger than the corresponding to $J_3$, thus governing the energy change in the spin wave spectrum.

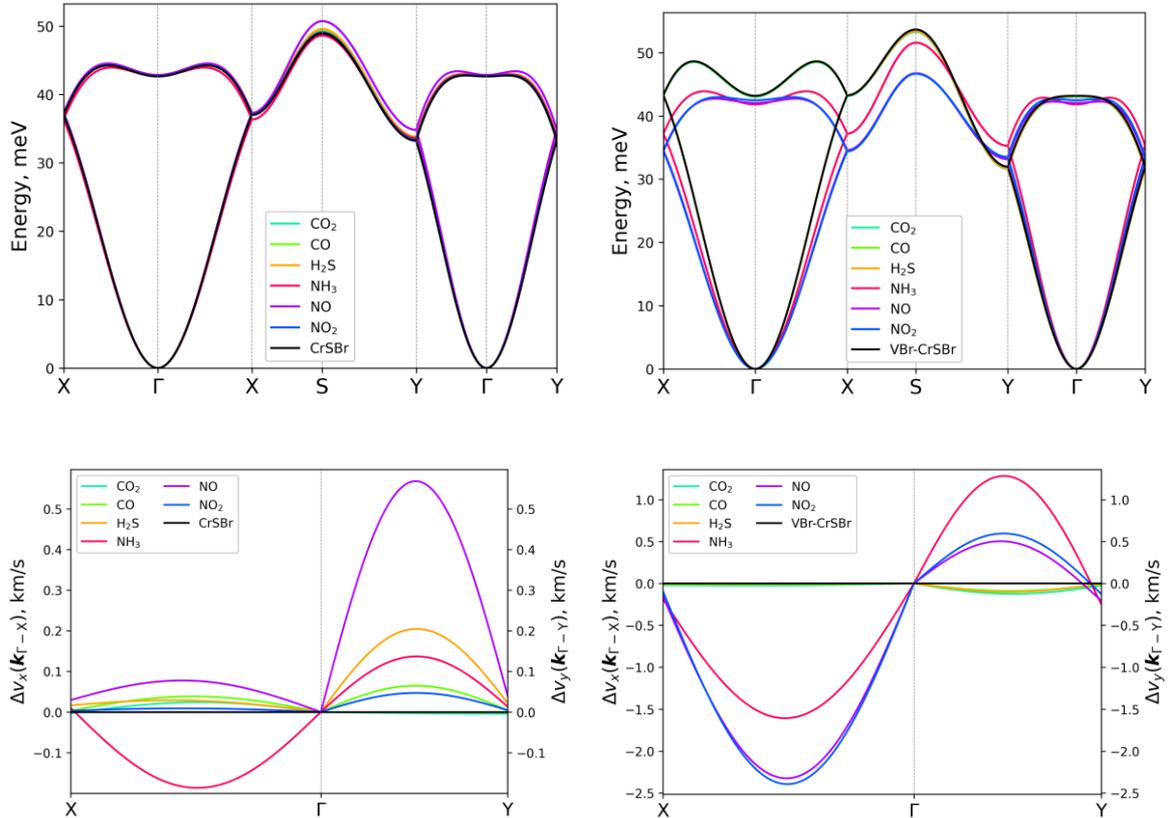

**Figure 5**. Magnon dispersion and relative change of group velocities of (a) pristine CrSBr and the hybrid gas/CrSBr heterostructures and (b) $V_{Br}$-CrSBr and the hybrid gas/$V_{Br}$-CrSBr heterostructures. Relative changes of group velocities of (c) gas/CrSBr and (d) gas/$V_{Br}$-CrSBr heterostructures.

These shifts in the magnon dispersion are reflected in the modification of the propagation velocity of the spin waves (**Figure 5c,d**). We focus our attention on the bottom magnon branch around the Γ point. As we can see, the changes in the group velocity of the spin waves, propagating along the direction of crystallographic lattice vector *a* are the most pronounced ones. The selectivity of the velocity modifications allows one to detect the presence of NO or $NO_2$ gases or $NH_3$, however, the differentiation between NO and $NO_2$ is not feasible according to our model.

Our results indicate that the propagation of magnons in single-layer CrSBr is altered when Br vacancies are present due to an increase of the magnetic moments around the defects. This is further affected by the chemical adsorption of specific N containing gas species ($NH_3$, NO and $NO_2$). In this context, both acoustic and optical magnon branches have been experimentally accessible in different vdW magnets including CrSBr[20] and other families such as $CrX_3$ (X= Cl, Br, I)[32,33] and $MPS_3$[34,35] (M= Fe, Mn, Ni). Hence, these materials and their defected structures may exhibit gas capturing or sensing behaviour according to the strength of the interaction between the substrates and the gas molecules, whose selective change in the magnons' frequency could be detected. We envision a new concept for gas sensing and capturing based on magnon measurements in 2D magnetic materials as an alternative to electronic ones, as far as a transducer is added to convert the magnetic response into an electrical signal that can be processed and analysed.

**Conclusions**

In conclusion, we have studied the adsorption behaviour of $H_2S$, $NH_3$, NO, $NO_2$, CO and $CO_2$ gas molecules over pristine CrSBr and the defected substrates $V_{Br}$-CrSBr and $V_S$-CrSBr by means of first-principles calculations. Our results show that the recovery time of pristine CrSBr for the adsorption of $CO_2$ is ~ 2.5 minutes, suggesting the possibility of using this substrate for its monitoring in the environment. In addition, we observe that the adsorption of $NH_3$, NO and $NO_2$ is really enhanced when a Br vacancy is present, offering the possibility to utilize VBr-CrSBr monolayer for capturing them. The impact of such molecules on the magnetic properties is noticeable, thus having a significant effect on the spin wave spectra and group velocities of magnons (~10%). Hence, we propose a new concept for gas sensing and capturing based on magnon measurements in 2D magnetic materials as an alternative to electronic ones.

**Computational details**

We performed spin polarized density-functional theory (DFT) calculations on single-layer CrSBr using the Quantum ESPRESSO package.[36] We used the generalized gradient approximation (GGA) as the exchange-correlation functional in the formalism of Perdew-Burke-Ernzerhof (PBE).[37] To take into account the electron correlation of 3$d$ orbitals of Cr, we selected an effective Hubbard $U_{eff}$ = 3 eV in the formalism of Dudarev et al.[38] The atomic coordinates and lattice parameters were fully optimized using the Broyden-Fletcher-Goldfarb-Shanno (BFGS) algorithm[39] until the forces on each atom were smaller than $1\cdot10^{-3}$ Ry/au and the energy difference between two consecutive relaxation steps was less than $1\cdot10^{-4}$ Ry. The electronic wave functions were expanded with well-converged kinetic energy cut-offs for the wave functions and charge density of 60 and 600 Ry, respectively. A vacuum spacing of 18 Å was set along $c$ direction to avoid unphysical interlayer interactions. The Brillouin zone was sampled by a Γ-centered 8×8×1 k-point Monkhorst–Pack grid.[40]

We constructed 4×4 supercells for the gas/CrSBr heterostructures to guarantee a minimum distance of 10 Å between the neighbouring gas molecules, thus avoiding interactions between them. In this case the Brillouin zone was sampled by a fine Γ-centered 4×4×1 k-point grid. The adsorption energy, $E_{ads}$, was calculated as:

$$E_{ads} = E_{CrSBr+molecule} - (E_{CrSBr} + E_{molecule}) \qquad (1)$$

where $E_{CrSBr+molecule}$, $E_{CrSBr}$ and $E_{molecule}$ are the total energies of the hybrid heterostructure, pristine CrSBr and the isolated gas molecule, respectively.

Charge transfer was calculated via Bader charge population analysis.[41] Also, we estimated the CDD by subtracting the charge densities of the isolated components (molecules and pristine CrSBr monolayer) from the total charge density of the combined CrSBr + gas molecule system by applying the following formula:

$$\Delta\rho = \rho\,(CrSBr + molecule) - \rho\,(CrSBr) - \rho\,(molecule) \qquad (2)$$

where ρ (CrSBr + molecule) and ρ (CrSBr) represent the total electron densities of CrSBr monolayer with and without adsorbed molecule, respectively, and ρ (molecule) represents the electron density of the isolated gas molecule. It should be noted that the separate molecule and substrate must have the same distorted geometry as in the hybrid system.

According to the transition state theory, the recovery time is calculated as follows:

$$\tau = \nu_0^{-1} e^{-\frac{E_{ads}}{kT}} \qquad (3)$$

where $\nu_0$ is the attempt frequency, k the Boltzmann constant, and T is the temperature (298 K). We assumed that the order of magnitude of $\nu_0$ for the tested gas molecules is similar as that of $NO_2$ ($\nu_0 = 10^{12}$ s$^{-1}$),[42] which it has been also considered in different works.[43,44]

The three magnetic exchange interactions ($J_1$, $J_2$ and $J_3$) were obtained by mapping total energies from different magnetic configurations on a classical Heisenberg Hamiltonian. In this case we considered $U_{eff}$ = 3 eV (U = 4 eV and $J_H$ = 1 eV, where $U_{eff}$ = U - $J_H$), showing results in good agreement with previous works.[45] Magnon dynamics is studied under the linear approximation of the Holstein-Primakoff boson expansion.[31] Details are described in the SI.